\documentclass[twocolumn]{aa} 
\usepackage{graphicx}
\usepackage{natbib}
\usepackage{txfonts}
\begin{document}

\newcommand{\fe}{[\ion{Fe}{ii}]}
\newcommand{\s}{[\ion{S}{ii}]}
\newcommand{\oi}{[\ion{O}{i}]}
\newcommand{\Ha}{H$\alpha$}
\newcommand{\pag}{Pa$\gamma$}
\newcommand{\pab}{Pa$\beta$}
\newcommand{\kms}{km\,s$^{-1}$}
\newcommand{\um}{$\mu$m}
\newcommand{\lam}{$\lambda$}
\newcommand{\ergscm}{erg\,s$^{-1}$\,cm$^{-2}$}

   \title{Tracing the origins of permitted emission lines in RU Lupi \\
down to AU scales\thanks{Based on observations collected  at the European 
Southern Observatory, Chile (ESO Programme 71.C-0703).}}

   \author{L. Podio\inst{1}
          \and
          P. J. V. Garcia\inst{2,3}
          \and
          F. Bacciotti\inst{1}
	  \and
	  S. Antoniucci\inst{4}
	  \and
	  B. Nisini\inst{4}
	  \and
	  C. Dougados\inst{5}
	  \and
	  M. Takami\inst{6}
          }

   \offprints{L. Podio}

   \institute{
     INAF-Osservatorio Astrofisico di Arcetri,
     Largo E. Fermi 5, I-50125 Firenze, Italy\\
     \email{lindapod@arcetri.astro.it, fran@arcetri.astro.it}
     \and
     Centro de Astrof\'{\i}sica da Universidade do Porto,
     Rua das Estrelas s/n, P-4150-762 Porto, Portugal\\
     \email{pgarcia@astro.up.pt}
     \and
     Departamento de Engenharia F\'{\i}sica,
     Faculdade de Engenharia da Universidade do Porto, Portugal
     \and
     INAF-Osservatorio Astronomico di Roma,
     Via di Frascati 33, I-00040 Monte Porzio Catone, Italy\\
     \email{nisini@mporzio.astro.it, antoniucci@mporzio.astro.it}
     \and
     Laboratoire d'Astrophysique de Grenoble, BP 53,
     38041 Grenoble Cedex, France\\
     \email{dougados@obs.ujf-grenoble.fr}
     \and
     Institute of Astronomy and Astrophysics, Academia Sinica, 
     P.O. Box 23-141, Taipei 10617, Taiwan, R.O.C.\\ 
     \email{hiro@asiaa.sinica.edu.tw}
   }

   \date{}

  \abstract
   {Most  of the observed emission lines and  continuum excess from
   young accreting low  mass stars (Classical T Tauri  stars -- CTTSs)
   take place in  the star-disk or inner disk  region.  These regions
   have a complex emission topology still largely unknown.}
   {In this  paper the magnetospheric  accretion and inner wind
   contributions to the observed permitted He and H near infrared (NIR) lines 
   of the bright southern CTTS RU Lupi are investigated for the first time.}
   {Previous optical observations of RU Lupi showed a large 
   H$\alpha$ profile, due to the emission from a wind in the line wings, 
   and a micro-jet detected in forbidden lines.
   We extend this analysis to NIR lines through seeing-limited 
   high  spectral resolution spectra taken with VLT/ISAAC, and  
   adaptive optics (AO) aided narrow-band imaging and low spectral resolution 
   spectroscopy with VLT/NACO. Using spectro-astrometric analysis
   we investigate the presence of extended emission down to very
   low spatial scales (a few AU).}
   {The HeI \lam10830 line presents a P Cygni profile whose 
   absorption feature indicates the presence of an inner stellar wind. 
   Moreover the spectro-astrometric analysis evidences the presence of an
   extended emission superimposed to the absorption feature
   and likely coming from the micro-jet detected in the optical.
   On the contrary, the origin of the Hydrogen Paschen and Brackett lines 
   is difficult to address. We tried tentatively to explain the observed
   line profiles and flux ratios with both accretion and wind 
   models showing the limits of both approaches. 
   The lack of spectro-astrometric signal indicates that the HI emission is
   either compact or symmetric. 
   Our analysis confirms the sensitivity of the HeI line to the presence of
   faint extended emission regions in the close proximity of the star.}
   {}

   \keywords{stars: individual: RU Lupi --
                stars: winds, outflows --
                accretion, accretion disks --
		line: formation --
		techniques: high angular resolution
               }

   \maketitle

\section{Introduction}

Classical T~Tauri stars  (CTTSs) are young solar and  lower mass stars
presenting  a   rich  emission line  spectrum  in  excess   of the 
continuum from the  stellar photosphere.    
This  emission   includes   Hydrogen
 lines, low energy forbidden lines and Helium
lines  together  with a  continuum  excess  --  the veiling.  

Optically and NIR  forbidden emission lines, such as the
[\ion{O}{i}], [\ion{S}{ii}] and  [\ion{Fe}{ii}] lines, have emission regions
large enough  to be resolved  and are typically associated with
outflowing jets \citep[e.g.][]{bacciotti00,dougados00,pyo06}.
On the other hand, the atomic permitted lines,  such as the Hydrogen lines,
are more difficult to  interpret. 
The energy released by the accretion of matter onto the star can power the
permitted line emission that is believed to arise
from magnetospheric    accretion    columns    and/or   accretion    shocks
\citep[e.g.][]{muzerolle01,bouvier07}.    The
fact  that  the  emission  component  of  permitted  atomic  lines  is
attributed to infalling gas is supported by the presence of inverse P~Cygni
profiles, and by the correlation of their luminosities  with  the accretion 
luminosity,
derived from the veiling \citep{muzerolle98b,muzerolle98c}.
On   the   other  hand,   the
blueshifted absorption features, when present, are an indicator of
high  velocity  inner  winds \citep[e.g.][]{edwards87,najita00}.   
Indeed, several recent spectroscopic studies suggest that  the emission of
the permitted HI and He 
lines  can actually include a contribution from both inflowing
and outflowing gas 
\citep[e.g.][]{beristain01,folha01,takami01,whelan04,edwards06}.
This has motivated hybrid models where the Hydrogen emission is due to
both the magnetospheric accretion 
 and a wind \citep{alencar05,kurosawa06}.

The fact that the collimated, spatially resolved jets, traced at large
distance from the source by optical, near-infrared forbidden lines, 
are generated in circumstellar regions (less than 10 AU from the source)
is predicted by theoretical models \citep{anderson03,ferreira06} and
observationally proved by the frequent detection of blueshifted absorption 
features in a number of strong permitted lines (H$\alpha$, Na, D, CaII, MgII, 
\citealt{najita00}), formed close to the star.
Signatures of outflowing gas revealed by absorption features can also be 
searched in the profile of the HeI \lam10830 line \citep{takami02,
edwards03,dupree05}.
\citet{edwards06} analysed a sample of 31 T Tauri stars and showed that the 
HeI line is much more sensitive than the usually used \Ha\, line to trace 
inner 
winds in absorption (in 71\% of the sources was detected an absorption feature 
in the HeI line against only 10\% of detections in the \Ha\, line).
\citet{beristain01} showed that the diagnostic power of permitted Helium lines 
in tracing inner winds, lies in their high excitation potential (20-50 eV), 
that restricts the line formation to a region of either high temperature or 
close proximity to a source of ionising radiation.\\

In order to investigate the reliability of such scenarios, we have
observed the circumstellar region of the T Tauri star RU Lupi in the
NIR wavelength range. 

This source  is one of  the most active  and well known T  Tauri stars
\citep{lamzin96,stempels02,herczeg05}  and  it  is  a good  target  to
investigate the  inner winds  and accretion flows  down to  very small
spatial  scales.  RU  Lupi is  located at  a distance  of only  140 pc
\citep{dezeeuw99},  allowing us  to reach  small spatial scales 
with high angular resolution instrumentation and/or specialised techniques
of data analysis.
In particular, NACO at the  VLT was used to take high  angular resolution 
images and low spectral resolution spectra.  
With VLT/ISAAC seeing limited and high spectral resolution spectra were 
obtained. A spectro-astrometric analysis was applied.
Spectro-astrometry is a powerful technique to obtain positional information 
on the region originating the emission in the components of a line profile
even in seeing-limited conditions.
\citet{takami01}, for example, using spectro-astrometry, found  that the
blueshifted and redshifted  wings of the H$\alpha$ line  emitted by RU
Lupi show a spatially symmetric offsets with respect to the source, in
the same direction of the jet traced, on higher spatial scales, by the
forbidden \oi\,  and \s\,  lines.  \citet{whelan04} detected  a similar
extended emission  in the wings  of the \pab\,  line in three  T Tauri
sources (DG Tau, V536 Aql, and LkH$\alpha$ 321).

In this  paper we  present the results obtained from such a study and,
in particular, a detailed analysis about the origin of the  Paschen and 
Brackett HI lines and the HeI \lam10830 line.
The paper is structured as follows: 
in Sect.~\ref{sect:data_reduction} we present the different set of
observations and the data reduction process;
in Sect.~\ref{sect:results} the observed line profiles and fluxes
are analysed; moreover we present the concepts underlying the 
spectro-astrometry technique and the obtained position spectra;
finally in   Sect.~\ref{sect:discussion} we discuss the origin of 
permitted H and He lines by means of a model for wind and accretion,
and of the results obtained with spectro-astrometry.
In the Conclusions (Sect.~\ref{sect:conclusions}) we summarize our findings.

\section{Observations \& Data Reduction}
\label{sect:data_reduction}

\begin{table*}[!ht]
\caption[]{\label{tab:obs} Instrumental setup during observations}
\begin{tabular}[h]{cccccccc}
\hline
Instrument & $''$/pix & Slit & Date & \lam\, (\um) & R & PA & T$_{int}$$\times$N$_{ex}$ \\ 
\hline
NACO    & 0.013 &            n.a.              & 20/25-5-03
& NB\,1.08,\,1.09 & n.a. & n.a. & 10.8s\,$\times$\,12 \\
imaging &       &                              &
& NB\,1.24,\,1.26 & & & 9s\,$\times$\,12    \\
     &       &                              &
& NB\,1.64,\,1.75 & & & 9s\,$\times$\,12    \\
\hline 
NACO & 0.054 & 0\farcs086\,$\times$\,40$''$ & 16-6-2003 & J: 0.91-1.37 & 400   & 217$^{\circ}$,\,127$^{\circ}$ & 18s\,$\times$\,10 \\
spectroscopy     &               &          & 22-4-2003 & H: 1.50-1.84 & 1500  &      &          \\ 
\hline
ISAAC & 0.147 & 0\farcs3\,$\times$\,120$''$  & 26-6-2003 & SZ: 1.06-1.1 & 11500 & 217$^{\circ}$,\,127$^{\circ}$ & 300s\,$\times$\,4 \\
      &       &                              &           & J: 1.24-1.3  & 10500 & 217$^{\circ}$,\,127$^{\circ}$ & 300s\,$\times$\,4 \\ 
      &       &                              &           & H: 1.54-1.62 & 10000 & 217$^{\circ}$                & 250s\,$\times$\,2 \\ 
\hline 
    \end{tabular}
\end{table*}

Observations of RU Lupi were carried out in between April and June 2003 at
the ESO Very Large Telescope. 
The first set of spectra was acquired at the UT4-YEPUN telescope using the 
infrared camera and spectrometer CONICA, working in cascade to the adaptive 
optics (AO) system NAOS, and the 86 mas width slit.
The AO system allows one to obtain a high spatial resolution of 
$\sim$0\farcs08-0\farcs12 (full width at half maximum - FWHM - 
of a point source), sampled by a pixel scale of 0\farcs054.
The second set of data was taken at the UT1-ANTU telescope using ISAAC
with the 0\farcs3 slit to obtain a high spectral resolution (R$\sim$10000;
spectral sampling: $\sim$14 \kms; 
pixel scale: 0\farcs147/pixel), while the spatial resolution is limited by the
seeing (FWHM$\sim$0\farcs6-0\farcs8). 
The NACO observations cover all the J and H bands wavelength ranges 
(R$\sim$400 and R$\sim$1500 respectively) while the ISAAC spectra centered at 
1.08, 1.27 and 1.588 $\mu$m cover very short wavelength segments. 
Both the NACO and the ISAAC spectra were obtained at two slit position 
angles (217$^{\circ}$ and 
127$^{\circ}$), i.e. parallel and perpendicular to the jet direction as 
determined by \citet{takami01}.

In addition two sets of images of the source in the narrow band filters 
centered around the HeI\lam10830 and [FeII]\lam1.25,1.64$\mu$m lines, and 
around their near continuum (at 1.09, 1.24 and 1.75 $\mu$m), were taken with 
NAOS/CONICA. 

The information about the spectral observations and the images are summarized 
in Tab.\ref{tab:obs}.\\


The spectra were reduced using standard IRAF tasks and Eclipse routines 
from the ESO pipeline.
The images were corrected for bad pixels and cosmic rays and for ghosts, 
then flat-fielded, and sky-subtracted. 
Wavelength calibration was performed using the spectra of the arc lamps for 
the NACO data and the sky OH lines for the ISAAC spectra.
The systemic motion of the object was subtracted using the radial velocity in 
the local standard of rest frame, determined  by \citet{takami01} through the
Li \lam6707.815 absorption line on the stellar atmosphere, 
V$_{LSR}$=8$\pm$2 \kms. 
The spectra were corrected for the atmospheric 
transmission and for the efficiency of the camera,
and then flux calibrated using a telluric spectro-photometric standard.
The typical uncertainty for the flux calibration is of $\sim$5\%.
Since the observed telluric stars are of type B, the absorption HI lines 
were subtracted before the correction.
Unfortunately the Br10, Br13 and Br15 absorption lines were deeply blended 
with the atmospheric features in the telluric spectra.
In the NACO spectra, because of the low resolution, it was not possible to 
perform a de-blending and subtract these HI absorption lines in the
spectrum of the telluric. 
Thus the  fluxes  and the equivalent widths of the Br10, Br13 and Br15 
lines measured from the NACO spectra are only upper limits.
In the ISAAC telluric spectra, on the contrary, the HI lines were de-blended 
from the atmospheric features and subtracted. 
Even at the higher spectral resolution of the ISAAC data, however, this process
is not very accurate and implies an oversubtraction.  
For this reason the fluxes, the FWHM and the equivalent widths 
of the Br10, Br13 and Br15 lines 
inferred from the ISAAC spectra should be considered as lower limits.

The NACO images were reduced using standard procedures: removal of bad 
pixel and cosmic rays, dark subtraction, flat-fielding and sky-subtraction. 
Then the images acquired with the narrow band filters centered on the 
continuum emission were subtracted from the images filtered on the line 
emission (HeI and \fe\,). The subtraction was performed by:
(i) choosing images with similar PSF, to minimize the effects of a different 
distribution of the emission in the different filters;
(ii) normalizing the images for the source flux estimated trough a gaussian 
bidimensional fit.

\section{Results}
\label{sect:results}

Analysis of  the NACO imaging  data showed that  
no physical  companion  is  detected  in the  continuum
frames in the limit of our spatial resolution ($\sim$ 14 AU).  
No extended emission  structure is detected in continuum
subtracted   emission    line   frames   at    HeI and
[\ion{Fe}{ii}]$\lambda$1.25,1.64$\mu$m.
This constrains the presence of wind emission in the HeI and [\ion{Fe}{ii}] 
lines over a region smaller than 14 AU.

\begin{figure*}[!ht]
   \centering
   \includegraphics[width=18.cm]{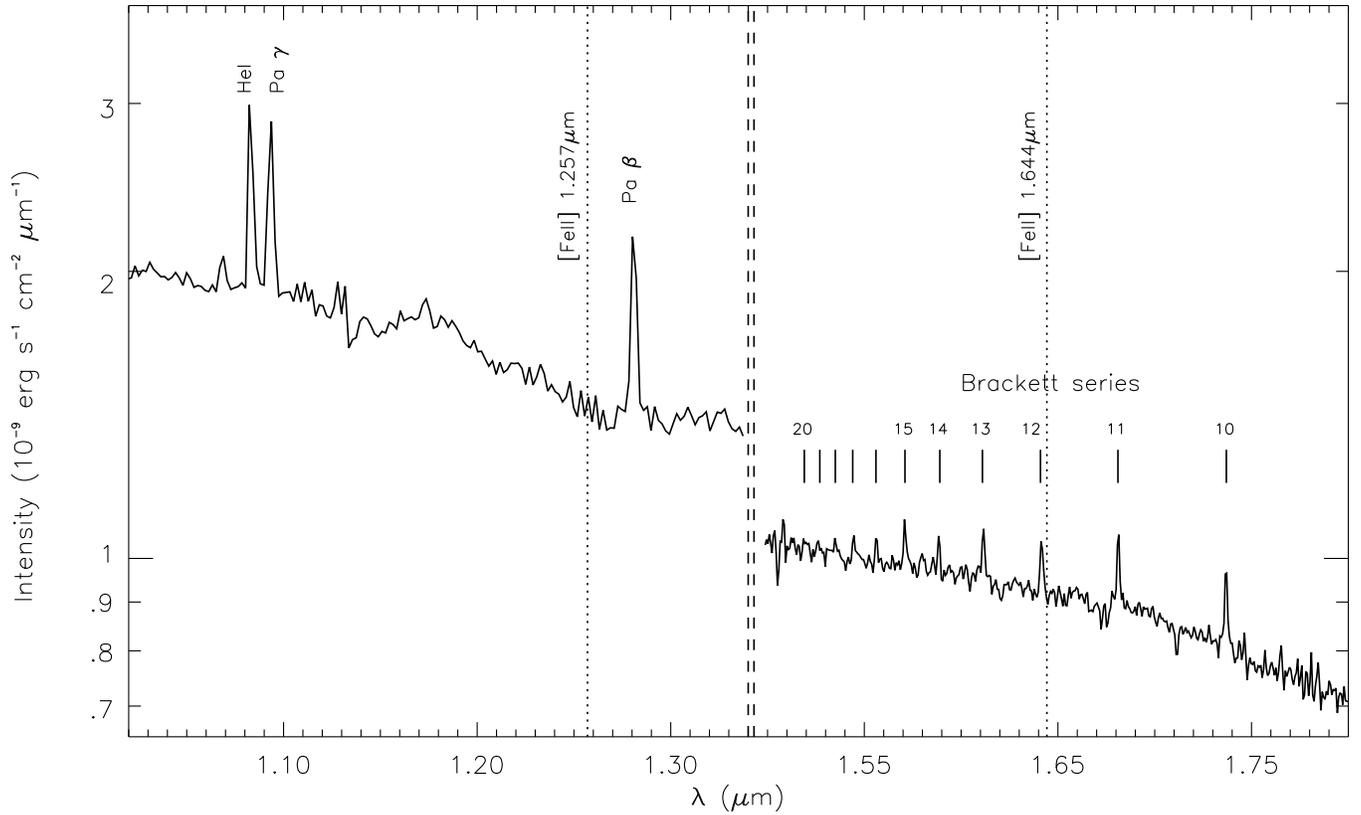}
      \caption{Low resolution NACO spectrum of RU Lupi in the J and H bands. 
    The HeI \lam10830 line and the HI emission lines from the Paschen and 
    the Brackett series are showed. 
    The dotted lines indicate the wavelength position of the \fe\, lines at 
    1.25 and 1.64\um, that were not detected in our spectra.}
         \label{spec_JH_217}
   \end{figure*}

\begin{figure}[!t]
   \centering
   \includegraphics[width=4.cm]{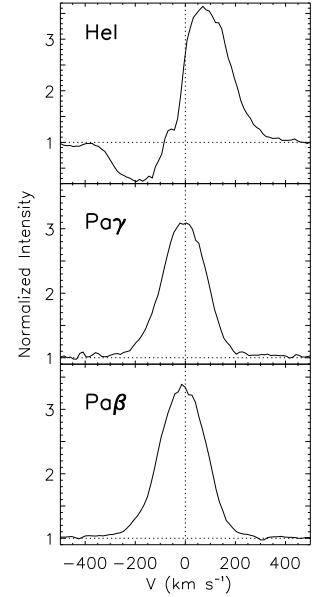}
      \caption{ISAAC spectra of the HeI \lam10830, \pag\,
      and \pab\, lines. The lines are normalized to the continuum intensity.}
         \label{isaac_spec}
   \end{figure}

The  NACO  low  resolution  spectra  show  strong  hydrogen
recombination  lines   in  emission  (Br10  to   Br19  and  Pa$\beta$,
Pa$\gamma$)  and the HeI 1.0830\um\, line (see Fig.\ref{spec_JH_217}
and Tab.~\ref{tab:NACO}).   
The  ISAAC  medium  spectral   resolution  spectra allow us to study in more
detail the spectral profiles of the HeI,  Pa$\beta$, Pa$\gamma$ and Br 13, 14,
15, 16 lines (see Fig.\ref{isaac_spec} and Tab.~\ref{tab:ISAAC}), 
as well as their spatial displacement 
with respect to the continuum emission through spectro-astrometry
(see Sect.~\ref{sect:spectro_astrometry}).  
No  [\ion{Fe}{ii}] 
lines are detected at  1.25, 1.32, 1.53, 1.60  and 1.64 $\mu$m both in the NACO
and in the ISAAC spectra.
The lack of [\ion{Fe}{ii}] emission is not expected in this source.
In fact, studies of HH jets in the optical and in the NIR range \citep{podio06}
showed that optical jets emit also in the NIR [\ion{Fe}{ii}] lines,
unless the density in the jet beam is very low (n$_{e}$ $<$ 10$^3$ cm$^{-3}$).
In RU Lupi, indeed, on one hand strong and spatially extended emission in the 
optical forbidden lines ([\ion{S}{ii}], [\ion{O}{i}], and [\ion{N}{ii}])
was observed by \citet{takami01};
on the other hand, electron densities higher than 10$^4$-10$^5$ cm$^{-3}$ were
estimated by the same authors based on the [\ion{S}{ii}] lines
equivalent width ratio.
Therefore, iron lines should be present in the spectra.
Possible reasons for the non detection of Fe emission could be the
fact that a short integration time was used, or that a strong depletion of iron
caused an emission fainter than the predicted one.
In fact, an increase of depletion near to the source has been observed in 
other jet sources as, e.g. HH 1 \citep{nisini05}.

\begin{table}[!ht]
\caption[]{\label{tab:NACO} Equivalent widths (EW) and dereddened 
line fluxes (F) from NACO spectroscopy.}
\begin{tabular}[h]{cccc}
\hline                                                                      
\lam   & Line   & EW   & F                  \\ 
\um    &        & \kms & 10$^{-13}$ \ergscm \\ 
\hline                                                                      
1.0830  &  HeI    & 452 &  40  $\pm$ 5          \\ 
1.0941  &  \pag   & 457 &  40  $\pm$ 5          \\ 
1.2822  &  \pab   & 434 &  32  $\pm$ 3          \\ 
1.5265  &  Br 19  &   9 &  0.5 $\pm$ 0.2        \\ 
1.5346  &  Br 18  &  14 &  0.8 $\pm$ 0.2        \\ 
1.5443  &  Br 17  &  20 &  1.1 $\pm$ 0.2        \\ 
1.5561  &  Br 16  &  19 &  1.1 $\pm$ 0.1        \\ 
1.5705  &  Br 15  &  46 &  2.7 $\pm$ 0.3  $^*$  \\ 
1.5885  &  Br 14  &  31 &  1.8 $\pm$ 0.2        \\ 
1.6114  &  Br 13  &  55 &  3.1 $\pm$ 0.3  $^*$  \\ 
1.6412  &  Br 12  &  48 &  2.7 $\pm$ 0.3        \\ 
1.6811  &  Br 11  &  63 &  3.5 $\pm$ 0.3        \\ 
1.7367  &  Br 10  &  78 &  3.9 $\pm$ 0.3  $^*$  \\ 
\hline                                                                       
\end{tabular}

$^*$ The equivalent widths (EW) and the fluxes (F) of the Br10, Br13 and Br15 line are only upper
limits to the real flux, due to the presence of HI lines blended with 
atmospheric features in the spectrum of the telluric standard
(see Sect.~\ref{sect:data_reduction}).
\end{table}

\begin{table}[!t]
\caption[]{\label{tab:ISAAC} Line properties from ISAAC spectroscopy.}
\begin{tabular}[h]{cccccc}
\hline                                                                      
\lam   & Line   & V$_{peak}$ & $\Delta$V & EW   & F$^a$              \\ 
\um    &        & \kms       & \kms      & \kms & 10$^{-13}$ \ergscm \\ 
\hline                                                                      
$^b$1.0830  &  HeI ab &      -200  &     -   & 133 &         -       \\ 
$^c$1.0830  &  HeI em &       70   &     -   & 581 &  47  $\pm$ 6    \\ 
1.0941      &  \pag   &       -5   &    206  & 475 &  40  $\pm$ 5    \\ 
1.2822      &  \pab   &      -11   &    214  & 562 &  38  $\pm$ 3    \\ 
1.5561      &  Br 16  &      -31   &    179  &  23 &  1.2 $\pm$ 0.2  \\ 
$^*$1.5705  &  Br 15  &      -16   &    128  &  20 &  1.0 $\pm$ 0.2  \\ 
1.5885      &  Br 14  &      -22   &    185  &  43 &  2.2 $\pm$ 0.2  \\ 
$^*$1.6114  &  Br 13  &      -19   &    159  &  44 &  2.7 $\pm$ 0.3  \\ 
\hline                                                                       
\end{tabular}

$^a$ The fluxes (F) are corrected for dust extinction 
(see Sect~\ref{sect:HI_lines}).\\
$^{b,c}$ Parameters relative to the absorption (ab) and emission (em) 
part of the HeI line profile (see Fig.~\ref{isaac_spec}).\\
$^*$ Due to the presence of HI lines blended with atmospheric features
in the spectrum of the telluric standard, the parameters inferred for the 
Br13 and Br15 lines are only rough estimates 
(see Sect.~\ref{sect:data_reduction}). 
In particular the width, $\Delta$V, and the flux, F, are lower limits.
\end{table}

\subsection{HI emission lines.}
\label{sect:HI_lines}

The availability of complementary data sets (see Tab.~\ref{tab:obs}) 
allows us to investigate the origin of HI lines through 
both the measured line fluxes and the line profiles.
The characteristics and the parameters of the lines profiles can be retrieved 
from the ISAAC spectra thanks to the high spectral resolution.
The centre (V$_{peak}$), the FWHM ($\Delta$V), the equivalent width (EW),
and the dereddened flux (F) of the HI 
lines which are in the spectral range covered by the ISAAC spectra (\pab, 
\pag\, and Br 13, 14, 15 and 16) are summarized in Tab.~\ref{tab:ISAAC}.   
The classification proposed by \citet{reipurth96} for the \Ha\, line profiles
was already used for NIR Hydrogen lines by \citet{folha01} and 
\citet{whelan04}.
According to this classification scheme, the detected HI lines
are type I profiles, i.e. they are nearly symmetric showing no evidence for 
absorption features, but they have large FWHM and
extended wings. 
All the Paschen and Brackett lines detected, in fact, 
are broad ($\Delta$V$\sim$200 \kms, 
except for the Br 13 and the Br 15 lines for which we get only a lower limit
to the FWHM) and slightly blueshifted ($-5 < V_{peak} < -30$).
In particular, the intensity profiles normalized to the continuum emission of 
the \pag\, and \pab\, lines are shown in Figure~\ref{isaac_spec}.  
Both the  \pag\, and \pab\, lines have a broad nearly symmetric profile 
peaking at 3-3.5 times the intensity of the continuum. 
The profile is centered at -5 and -11~\kms, respectively, 
and extends at least from -300~\kms  to  +200~\kms. 


   \begin{figure}[!t]
   \centering
   \resizebox{\hsize}{!}{\includegraphics{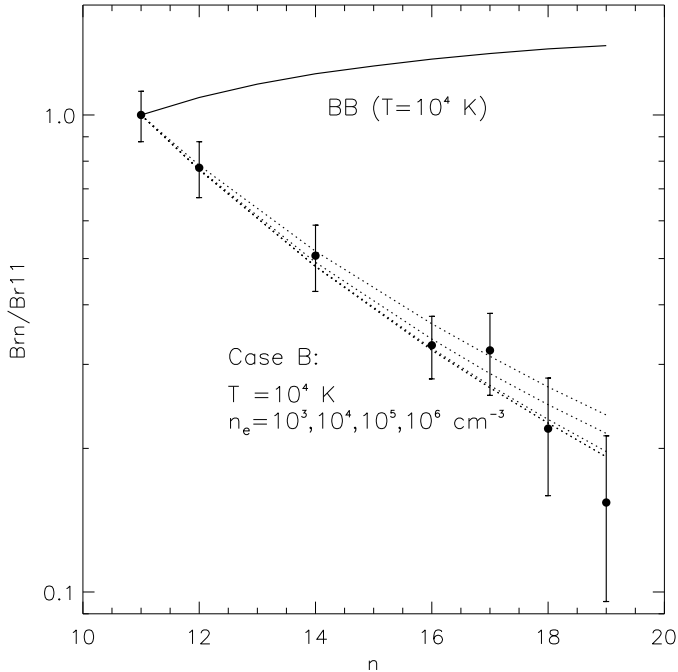}}
      \caption{Ratios of the Brackett lines with respect to Br11 versus
      the upper quantum number $n$. The solid line indicates the theoretical 
      value of the ratios expected
      in case of optically thick emission at T=10\,000 K. The four dotted lines
      are the ratios expected from case B recombination (Hummer \& Storey 
      1987) assuming T=10\,000 K and $n_e$=10$^3$, 10$^4$,  10$^5$, 10$^6$ 
      cm$^{-3}$ going from the lower to the upper line.}
         \label{br_br11}
   \end{figure}


The ratios between the lines of the Paschen and Brackett series can be used 
to retrieve further informations about the origin of the HI lines.
The NACO spectra have too low spectral resolution to derive the
properties of the line profiles (V$_{peak}$, $\Delta$V).
They can be used, however, to measure the equivalent widths and the fluxes of 
the Paschen and Brackett lines (see Tab.~\ref{tab:NACO}). 
With respect to the ISAAC data, these spectra have the advantage of covering 
all the Brackett lines from
the Br 10 up to the Br 19 in the same spectral segment, allowing us to measure
and compare the fluxes of all the lines without introducing calibration
problems.  
First, we corrected the observed fluxes for the reddening.
We used the average value of the visual extinction estimated by 
\citet{giovannelli95} in the circumstellar region of RU Lupi 
(A$_V$=0.7 $\pm$ 0.3) and the extinction law derived by \citet{rieke85}.
The dereddened line fluxes of the Paschen and the Brackett lines
are summarized in Tab.~\ref{tab:NACO} for the NACO data and in
Tab.~\ref{tab:ISAAC} for the ISAAC ones.
The error affecting the line fluxes and their ratios is computed considering 
the noise over the continuum and the uncertainty on the A$_V$ value.
Then we computed the Brackett decrement, i.e. the ratios between the intensity
of the Brackett lines coming from the upper level $n$ and the first line of the
series. The first Brackett line in the spectral range covered by our 
spectra is the Br 10, but this line is not well corrected for the atmospheric
transmission (see Sect.~\ref{sect:data_reduction}).
Thus we used the ratios with the Brackett 11 line.
The Brackett ratios, Br$n$/Br11, as a function of the upper level $n$, are 
plotted in Fig.~\ref{br_br11}
(the ratios Br15/Br11 and Br13/Br11 are not taken into consideration
since for the Br13 and Br15 lines we have only an upper limit of the
flux, see Sect.~\ref{sect:data_reduction}).
In the Figure we overplotted the theoretical values of the Brackett ratios 
calculated for two different cases:
(i) the solid line traces the ratios in the case of optically thick emission 
at the temperature T=10$^4$~K (blackbody emission);
(ii) the four dotted lines correspond to Case B recombination \citep{hummer87}
for T=10$^4$~K and four different values of the electron density.
Note that for the considered values of the electron density, variations of the
temperature between 10$^3$ and 10$^5$~K leave the curves almost identical.
The plot of Fig.~\ref{br_br11} shows that the ratios between the
Brackett lines are well fitted by case B recombination,  i.e. by optically
thin emission. 
As we will discuss in Sect.~\ref{sect:HI_model}, however, 
case B cannot reproduce the observed fluxes
for typical values of the emitting volume.

To avoid systematic errors that can affect the lines of the same series, and to
constrain even more tightly the emission conditions,
we considered also the ratio between the observed Paschen lines, \pab/\pag.
The measured \pab/\pag\, ratio retrieved from the NACO spectra
is $\sim$0.8$\pm$0.1, and the one retrieved from the ISAAC ones is 
$\sim$1$\pm$0.1.
Such values cannot be produced by the case B 
recombination. 
In fact, for any choice of the electron density and the
temperature between 10$^2$ and 10$^{10}$ cm$^{-3}$ and 5 10$^2$
and 3 10$^4$ K, respectively, case B gives \pab/\pag\,$>$1.4.
On the other hand, in the case of optically thick emission 
(blackbody emission) the \pab/\pag\, ratio is $\sim$0.6 at
T=10$^4$ K, and $\sim$0.7 at T=5 10$^3$ K.
This implies that the Paschen lines are not reproduced by one of these 
limiting cases 
and that their ratio strongly depends from their relative optical depth.  



\subsection{The HeI \lam10830 line.}
\label{sect:HeI}

The HeI line profile obtained from medium resolution ISAAC spectra 
(Fig.~\ref{isaac_spec}) 
shows a typical P-Cygni profile with the emission part peaking at  
+70~\kms, and extending up to at least +430~\kms, and a broad blueshifted 
absorption penetrating up to 80\% of the continuum at V$\sim$-200~\kms
and ranging from -360~\kms\,  to -90~\kms (see Tab.~\ref{tab:ISAAC}).
The absorption feature clearly indicates the presence 
of an inner wind \citep{edwards03}.
The efficiency of the HeI \lam10830 transition in tracing outflowing gas is 
due to the ``metastability'' of its lower level (2$s^{3}$S).
Even if energetically far above the singlet ground state (21 eV), this level
is radiatively isolated from it and collisionally de-excitation is very low.
This means that, once it is populated, via photoionisation
followed by recombination and cascade and/or via collisional excitation from
the ground state, this level becomes easily over-populated.
The absorption feature is generated mainly via resonant scattering 
since the \lam10830 transition is the only permitted radiative transition
from its upper state to a lower one and  the electron density 
is unlikely high enough to cause collisional excitation or de-excitation.
\citet{edwards03,edwards06} demonstrate that the shape and 
width of the HeI \lam10830 line can constrain the wind launch region.
They show that a broad and deep blue absorption, as the one detected in our 
spectra, require formation in a wind emerging radially from the star 
rather than from the disk.
In fact, the broad range of velocities covered by the absorption feature
can be explained by radially emerging wind streamlines absorbing the 
1 \um\, continuum from the stellar surface, thus tracing the
full acceleration region of the inner wind.
On the contrary, a disk- or x- wind is confined to nearly parallel streamlines
emerging at an angle to the disk surface and thus the continuum photons
from the star will intercept a narrower range of velocities giving rise
to narrow absorption feature as the ones shown in  \citet{edwards06}.
In conclusion, the absorption feature in our spectra appears to indicate
the presence of an inner wind radially emerging from the star.
Further insight on the geometry and the presence of different wind 
contributions to the HeI \lam10830 line
can be retrieved from the spectro-astrometric analysis presented in
Sect.~\ref{sect:spectro_astrometry} and discussed in 
Sect.~\ref{sect:spectro-astrometry_disc}.

The emission part of the line, instead, is produced by the accreting gas 
columns and, partially, by the resonant scattering process.
The contribution from the latter depends on the electron density
and in our case should be much lower than the loss of flux in the absorption 
part of the line.
In fact, the detection of the absorption feature indicates that the efficiency 
of collisional de-excitation is much lower with respect to the radiative decay.   


\subsection{Spectro-astrometry}
\label{sect:spectro_astrometry}

%
   \begin{figure*}[!t]
   \centering
   \includegraphics[width=\columnwidth]{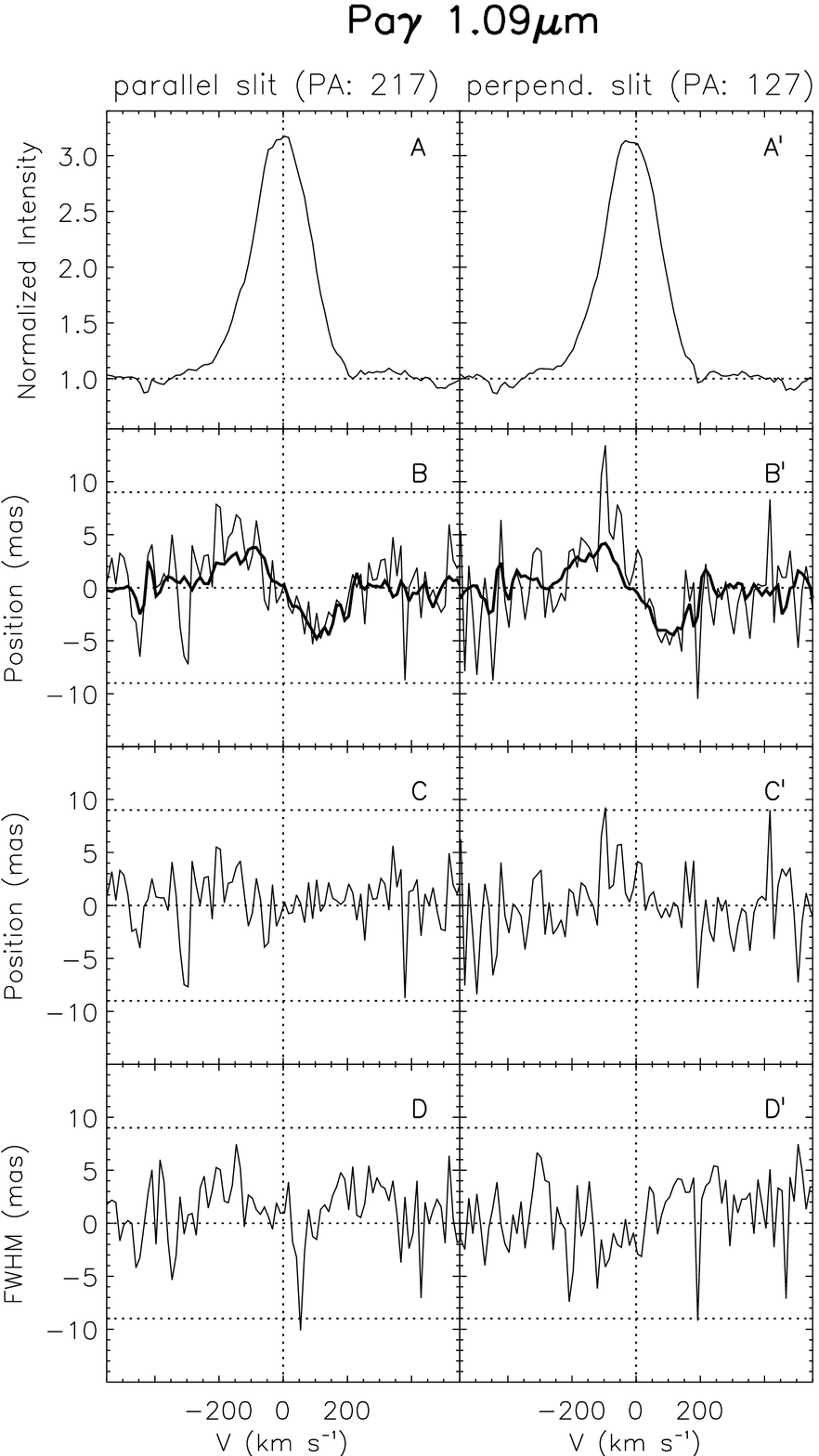}
   \includegraphics[width=\columnwidth]{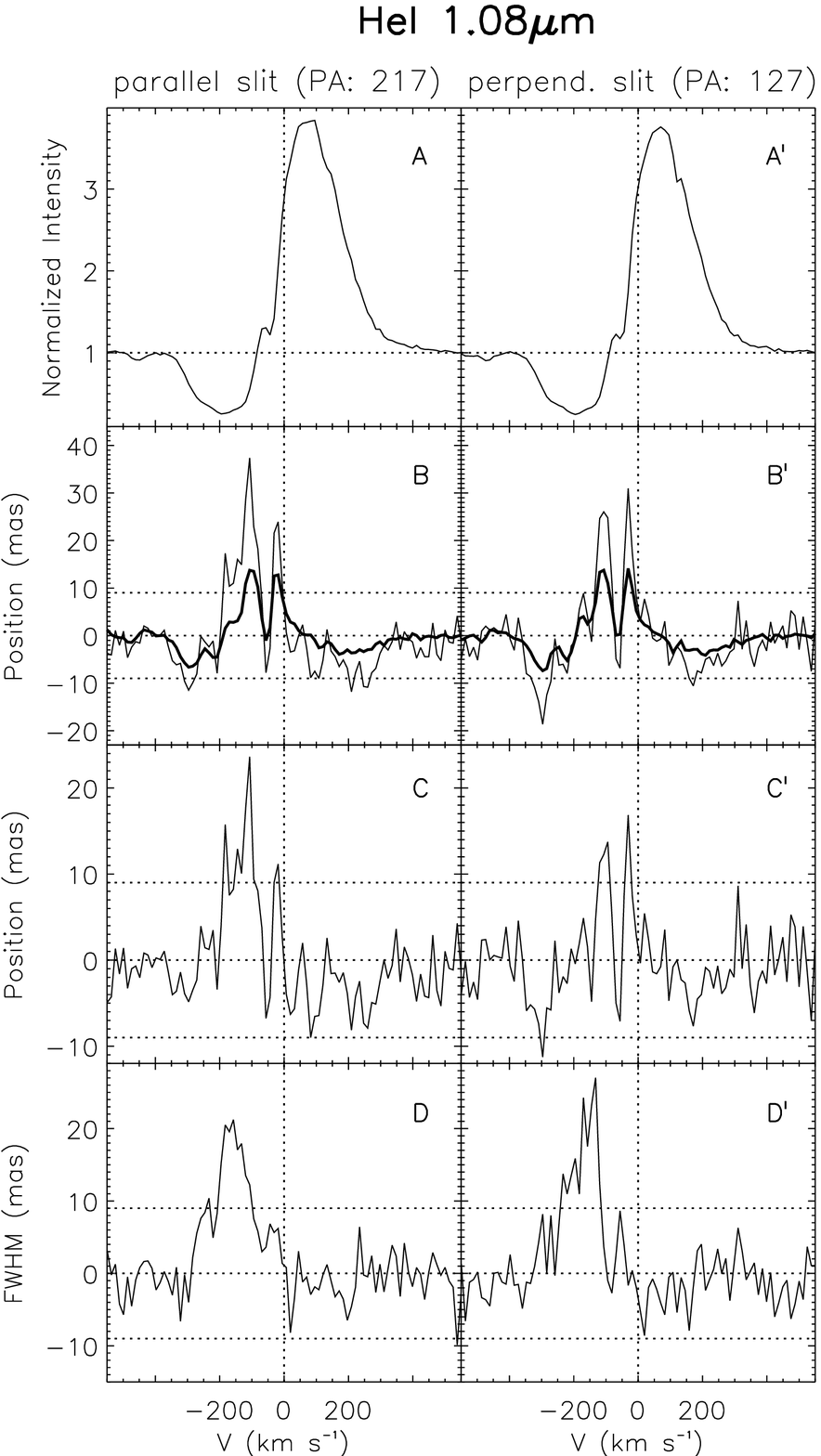}

      \caption{ISAAC spectroscopy of RU Lupi Pa$\gamma$ (left) and HeI
      (right), with  the slit parallel  to the jet  and perpendicular.
      The line profiles normalized to the continuum emission are
      presented   in   the   upper   panels   --  A   and   A'.    The
      spectro-astrometric signal is presented  in the second panels --
      B and  B'. The overplotted solid line is the spectro-astrometric
      signal due to a distorted PSF caused by tracking error or unstable
      active optics. The signal is simulated following the procedure
      of \citet{brannigan06} (see text). The residual signal obtained
      subtracting the simulated signal to the detected one is shown in
      the third panels  --  C   and   C'.
      Finally, the residual FWHM of the  gaussian fit is
      presented  in the lower  panels --  D and  D'. 
      The vertical dotted lines show the zero velocity of the line.
      The  horizontal dotted lines  indicate the continuum
      emission  level  and  the  $\pm3\sigma$  threshold  due  to  the
      signal-to-noise.   
      \label{spectro_astrometry}}
   \end{figure*}


Further  to the standard  spectroscopic analysis, we performed a 
spectro-astrometric analysis of the data.
Following  this procedure the  displacement of the emission centroid  
of the lines with respect to the source continuum is determined
\citep[e.g., ][]{bailey98}. 
This means that one can detect extended emission in a line beyond the seeing
limit, down  to  (sub-)AU  scales.
On the other hand, the technique is sensitive only to asymmetric emission and, 
since it measures a barycentre, it gives only a lower limit to the size of the 
emission region.
The technique consists in accurately measuring the trace of each A-B spectrum  
by Gaussian fitting  the spectrum  spatial profile  at each  wavelength. 
In this way one recovers the wavelength dependent  centroid (i.e. the so
called ``position spectrum'') and gaussian FWHM of the spectra.   
The residual FWHM, obtained subtracting the FWHM of the continuum emission, 
can be  sensitive to differential size variations  in the emission
region.
Examples of the application of spectro-astrometry can be found in 
\citet{takami01}, \citet{whelan05}.

The essential requirements for this kind of analysis are a
good spectral and spatial sampling, and a high signal-to-noise ratio (SNR). 
Thus we applied spectro-astrometry only to the ISAAC spectra, in which the
lines are  very well resolved (SNR$\sim$100) and well sampled spectrally  
($\Delta$V$\sim$200 \kms, spectral sampling $\sim$14 \kms), while the spatial
sampling is moderately good (FWHM$\sim$0\farcs6-0\farcs8, 
spatial sampling: 0\farcs147).
Since the trace of the star is tilted in the ISAAC spectrograph, the  
centroid  is  corrected  for  this distortion with a polynomial  fit.  
The typical spectro-astrometric error is given by the accuracy in measuring
the centroid of the spatial profile through the gaussian fit and is equal to 
$\sigma$$\sim$0.5\,\,FWHM/$\sqrt{N_{ph}}$ \citep{takami03}.
For the considered lines ($N_{ph}$$>$10$^{5}$), this
accuracy turns out to be $\sim$1 mas. 
In the case of the He line, however, the signal is fainter in the absorption
part ($N_{ph}$$\sim$10$^{4}$-10$^{5}$). 
Therefore, we adopted $\sigma$$\sim$3 mas, which is the average value of 
the error across the line profile.
Moreover, since the spatial sampling for the ISAAC data is not exceptional, 
we used a conservative value of the threshold of 3$\sigma$ 
(see Fig.~\ref{spectro_astrometry}).
We checked the position spectra for the presence of artifacts following 
the procedure of \citet{brannigan06}.
These authors showed that the distortion of the PSF caused by tracking 
errors of the telescope or unstable active optics during an exposure, 
can induce artificial signals.
These instrumental biases can be simulated in shape and magnitude 
calculating the spectro-astrometric
signal obtained from a modeled distorted PSF.
We used the same procedure than in \citet{brannigan06}
to simulate the biases that can be produced with our 
instrumental setting (cf. Tab.~\ref{tab:obs}).
The shape of the artifacts produced by a distorted PSF
depends on the slope of the line profile.
In particular, the peaks of the simulated biases appear
at velocities where the gradient of the line profile is the
largest. This explain the double peak of the
simulated spectro-astrometric signal in  the blueshifted part of the
HeI line (see panel B and B' of Fig.~\ref{spectro_astrometry}).
The magnitude of such biases, instead, depends on the amplitude
of the distortion of the PSF with respect to the slit axis, 
i.e. on the parameters 
$\Delta x_1$, $\Delta x_2$, $\Delta y_1$, $\Delta y_2$ in 
Eq. (4) of \citet{brannigan06}.
The maximum biases are obtained when the distortion of the PSF in the
slit is maximum, i.e. when we put in Eq. (4) of \citet{brannigan06}
$\Delta x_1= \Delta x_2=0\farcs15$, 
$\Delta y_1= \Delta y_2=0\farcs15$, so that 
the total slit illumination displacement ($\Delta y$) is equal to the slit 
width, 0\farcs3.
The resulting spurious signal was then subtracted from the position spectra.

The position spectra obtained for the \pab\, and Brackett lines do not show 
any shift of the centroid with respect to the continuum emission.
On the other hand, in the \pag\, and the HeI line we found a 
spectro-astrometric signal (see Fig.~\ref{spectro_astrometry}).
In panels A and A' the intensity spectrum of the two lines in the two
slit positions is shown.
The position spectrum, in panels B and B', indicate a shift of the centroid of 
the emission with respect to the continuum emission both in the parallel
and in the perpendicular slit.
The superimposed solid line is the spurious spectro-astrometric signal 
that can be induced by an elongated PSF in the slit in the worst possible case.
In panel C and C' the residual signal obtained subtracting the determined 
biases is shown.
Finally in panels  D and D' we plot the residual FWHM of the gaussian fit. 

Fig.~\ref{spectro_astrometry} shows that there is no real signal 
in the \pag\, line.
In fact, there is no residual in the position spectra of this line 
after the subtraction of the biases as well as 
in the  spatial  FWHM  across the  profile.
In the HeI line, on the contrary, the position spectra cleaned by artifacts
(panels C and C') show a shift over the signal-to-noise limit 
(9 mas, corresponding to $\sim$1.3 AU)
in the blueshifted part of the line.
The residual FWHM across the profile presents a significant signal  in the  
same velocity range  where the spectro-astrometric signal is detected 
(-50 \kms$<$V$<$-200 \kms), reaching differential sizes of $0\farcs025$.
A broader FWHM is expected in the region where the absorption
feature is detected, because the continuum emission is much fainter.
It is, however, interesting to notice that:
(i) the spatial profile can be identified very well over the background noise 
across the entire HeI spectral profile (the signal-to-noise ratio is
S/N$\sim$100 in the spectral region covered by the absorption feature
and $\sim$10 times higher in the emission part of the line); 
(ii) the broadening of the line is not symmetric with respect to the peak
of the absorption feature (V$_{peak}$$\sim$-200 \kms). Instead, a constant
or increasing FWHM in the parallel and perpendicular slit respectively, is
measured in the region where the spectro-astrometric signal is detected
(-200 \kms$<$V$<$-100 \kms).
This suggests  that   multiple   spatial  components   are
contributing  to the  line profile  in such region of the  ``P  Cygni'' 
absorption feature.
The interpretation of the detected spectro-astrometric signal is
discussed in Sect.~\ref{sect:spectro-astrometry_disc}.


\section{Discussion}
\label{sect:discussion}

As illustrated in the Introduction 
the excitation mechanism of the HI lines is an actively debated 
topic. 
Up to a few years ago, both the observations and the models focused 
on the Balmer lines, and in particular on the \Ha\, line.
The large amount of optical observations evidenced the limits of both the
accretion and the wind models in reproducing the Balmer lines and motivated
the development of hybrid models with more complex geometries 
\citep[e.g., ][]{kurosawa06}.  
On the other hand, the capacity of these models to reproduce the profiles
and fluxes of the HI infrared lines was poorly tested.
Interestingly, the observations of Paschen and Brackett lines, 
such as the ones presented in \citet{folha01}, \citet{nisini04},
\citet{whelan04}, \citet{edwards06},
have shown that the magnetospheric accretion models
cannot fully reproduce the characteristics of the observed profiles, in favour
again of an hybrid wind/accretion model.

As we have shown in the previous section, the Paschen and the Brackett lines
in our spectra present blue-shifted broad symmetric profiles
(V$_{peak} \sim$ -17 \kms and $\Delta$V $\sim$ 200 \kms).
These are difficult to reproduce if one adopts {\em separately} 
a wind model or, alternatively, a  magnetospheric
accretion model.
Wind models generally predict red-shifted peaks 
and strongly asymmetric 
profiles with P-Cygni absorption features, as the one detected in the HeI line 
\citep{calvet92}.
\citet{hartmann90} were able to reproduce more symmetric Br$\gamma$ and 
Br$\alpha$ profiles, without the P-Cygni absorption feature with a model of 
magnetically driven wind, but their peaks are still red-shifted.
\citet{edwards87} showed that an optically  thick circumstellar disk may cause
the occultation of the red-shifted part of the profile thus explaining
blue-shifted peaks. In any case, asymmetric profiles are expected.

On the other hand, magnetospheric accretion models \citep{muzerolle98a} 
predict blue-shifted peaks, but asymmetric and narrow profiles 
($\Delta$V $\sim$ 100 \kms) often showing inverse P-Cygni absorption.
\citet{muzerolle01} retrieved from their accretion model more symmetric and
broad line profiles, taking into account line damping due to different 
broadening mechanisms, but this effect is really important only for \Ha\, 
and Balmer lines, while it is negligible for Paschen and Brackett lines.

We will now examine how the above scenarios fit in the case of our RU Lupi
spectra.

\subsection{Accretion vs wind}
\label{sect:HI_model}

As illustrated in Sect.~\ref{sect:HI_lines} the case B recombination 
well reproduces the Brackett decrement, but not the \pab/\pag\, ratio,
whose value implies that these two lines are optically thick.
Moreover, case B recombination assumes 
(i) a constant density, which is not 
applicable to the circumstellar region of T Tauri stars, and
(ii) optically thin lines.
To reproduce the observed fluxes, however, the latter hypothesis would imply 
an emitting volume corresponding to a spherical envelope extending up to 40 AU 
from the source, which is unrealistic according to the models of circumstellar 
accretion/ejection structure \citep[e.g., ][]{nisini04}.
To investigate the origin of the HI lines one has to assume a variable density
profile, and the optical depth and the flux of the Paschen and
Brackett lines will be retrieved integrating along the line of sight. 

\citet{nisini04} analysed the Brackett lines in the Class I source HH\,100 
IR, finding line profiles very similar to the ones of our Paschen and
Brackett lines.
Interestingly, they found that the Brackett decrement in HH\,100 IR is neither
reproduced by the Case B recombination curves (i.e. optically thin
emission), nor by the blackbody emission (i.e. completely thick emission). 
The authors suggested that these lines are emitted in an 
expanding envelope of ionised Hydrogen, and thus different lines
are tracing different layers of the envelope according to their optical depth. 
Following this idea they were able to reproduce the ratios between the 
optically thick Brackett lines through a simple model of 
spherical wind.

We used the same approach to reproduce 
the Br$n$/Br11 ratios (Fig.~\ref{br_br11}), the \pab/\pag\, ratio
 and the observed line fluxes
both in the accretion and in the wind hypothesis adopting a toy 
model consisting in a spherical envelope of ionised Hydrogen, 
extending from an inner radius $r_i$ to an outer radius
$r_{out}$, where LTE conditions are assumed. 
If the envelope is expanding (wind hypothesis) or infalling 
(accretion hypothesis)
the radiation transfer can be treated in the
Sobolev approximation following the formalism described in 
\citet{castor70} and \citet{nisini95}. 
The gas can be only partially ionised, 
and, therefore, the envelope total mass loss or accretion rate is 
$\dot{M}_{wind/accr} = \dot{M}_{ion} / x_e$, where $ x_e$ is the ionisation
fraction.
The electron density in the envelope is given by the 
continuity equation, 
$n_e=\dot{M}_{wind/accr} x_e / 4 \pi m_H r^2 V(r)$.

Using this model both the wind and the accretion hypothesis can be tested
just changing the velocity law $V(r)$.
In fact, this will imply a different density profile and thus a difference in 
the relative optical depths of the lines in the two cases.

In the wind hypothesis the HI envelope is expanding and 
the gas velocity follows a general law of the type  
$V(r) = V_{0} + (V_{max} - V_{0}) (1 - (r_i/r)^{\alpha})$, i.e. the gas is
accelerated  from an initial velocity at the base of the wind, $V_{0}$,
at the maximum velocity $V_{max}$ 
at a distance that depends on the parameter $\alpha$ \citep{nisini04}.
V$_{max}$ is the maximum velocity of the expanding gas.
The FWHM of the observed lines is determined  by the 
velocity of the gas particles, thus we retrieve from the observed profiles
V$_{max}$$\sim$200 \kms.

The simulated emission is tuned by a proper choice of the input parameters.
For typical values of the size of the envelope ($r_i =1\, R_{*}$,
$r_{out} = 3-10\, R_{*}$) and of the electron temperature (T = 10$^4$ K), 
and taking  $\alpha$=4 and $V_{0}$=30 \kms as
in \citet{nisini04}, we are able to reproduce the 
Br$n$/Br11 ratios and the absolute fluxes of the Paschen and almost all 
the Brackett lines (within $\pm$4$\sigma$) 
for $\dot{M}_{ion} \sim 2\,10^{-8}$ M$_{\odot}$ yr$^{-1}$.
On the other hand, in this way we are not able to reproduce the observed 
\pab/\pag\, ratio (\pab/\pag$\sim$1.3-1.4).
Considering a slower acceleration of the wind ($\alpha$=1-3) and/or a 
lower temperature (T down to 5 10$^3$ K) the \pab/\pag\, ratio decreases to
$\sim$1-1.3 and the absolute fluxes are in agreement with the
observed ones but the Brackett ratios are not reproduced because the lines 
optical depths are too large.
Taking lower values of the mass loss rate 
($\dot{M}_{ion} < 10^{-8}$ M$_{\odot}$ yr$^{-1}$), that is the
fundamental parameter regulating the emission in the HI lines,
we obtain a good fit of the Brackett decrement for
$r_{out}$ within $3-10\, R_{*}$ and values of the exponent in the acceleration 
law of the wind between 1 and 4.
The absolute fluxes, however, are too low in this case (3-4 times
smaller than the observed ones) and the \pab/\pag\, ratio is $>$1.4.
To obtain the required \pab/\pag\, one should assume
a value of $\dot{M}_{ion}$ higher than  $2\,10^{-8}$ M$_{\odot}$ yr$^{-1}$.
For such a massive wind, however, the Brackett decrement and line fluxes
would no longer be fitted by the model because their optical depth would be
too large.

Since we could not find a choice of the parameters for the wind model 
reproducing 
at the same time the Brackett decrement and the observed \pab/\pag\,
ratio, we tentatively tried to use the  model to simulate the emission from 
a spherical accreting envelope, by reversing the sign of the gas speed.
A spherical model is not the best approximation of the
accretion process, that, according to the most recent models,
proceeds along the magnetic field lines, in an axisymmetric 
non-spherical geometry \citep{hartmann94,muzerolle98a}.
Nevertheless, we have adopted the spherical model  as a first approximation, 
in order to check if the observed ratios are better reproduced by accretion 
rather than by an outflow.

If we assume that the HI lines are mainly excited in the accretion columns
we can derive an estimate of the mass accretion rate from the luminosity
of the \pab\, line.
To this aim, we use the empirical relation between L(\pab) and the accretion
luminosity derived by \citet{muzerolle98c} and \citet{natta02}:
\begin{equation}
\log L_{acc}/L_{\odot} = 1.36 \log L(Pa \beta)/L_{\odot} + 4.
\end{equation} 
The mass accretion rate is then computed from L$_{acc}$ as
$\dot{M}_{acc} = L_{acc} R_{*} /(G M_{*})$.
Using for RU Lupi M$_{*}$ = 0.8 M$_{\odot}$ and R$_{*}$ = 1.6 R$_{\odot}$
\citep{lamzin96,herczeg05}, we obtain $\dot{M}_{acc}$=2\,10$^{-7}$ 
M$_{\odot}$ yr$^{-1}$.
This value is in agreement with previous estimates of \citet{lamzin96}, that
tuned the value of $\dot{M}_{acc}$ in order to reproduce through their
model the observed continuum energy distribution (see, however,  
\citealt{herczeg05}, for an alternative derivation of $\dot{M}_{acc}$ 
using the accretion luminosity estimated from the UV excess).

In the case of an accreting envelope the typical velocity law is 
$V(r) = V_{max} r^{-0.5}$. 
We fixed the size of the envelope and the temperature 
($r_i=R_{*}$, $r_{out}= 3 R_{*}$, T=10$^4$~K),
and examined the results
obtained varying the ionisation fraction  x$_e$, and thus the electron density
of the accreting columns for  $\dot{M}_{acc}$=2\,10$^{-7}$ 
M$_{\odot}$ yr$^{-1}$.
If the gas is fully ionised (x$_e$=1) we obtain
\pab/\pag$\sim$0.7, but the Brackett decrement is not reproduced,
because the optical depth of these lines is too large,
and the absolute fluxes of all the lines are higher than the observed one
by one order of magnitude.
Indeed, estimates of the ionisation fraction for different values of the
temperature and the mass loss rate show that the gas is only partially 
ionised in the circumstellar envelope \citep[e.g., ][]{natta88}.
Assuming x$_e$=0.4 
both the Brackett decrement and the ratio between the Paschen lines are well
reproduced, but the absolute fluxes are still too high by a factor 3-4.
To obtain absolute fluxes in agreement with the observed ones we have to assume
low ionisation fractions, down to x$_e$=0.15. 
In this case, however, we are not able to reproduce
the \pab/\pag\, ratio.
We note that 
in a non-spherical geometry the lines become optically thick for lower
mass accretion rates, and thus both a ``thick'' \pab/\pag\, ratio and lower 
absolute fluxes can be obtained at the same time.
In such a model, however, the profiles of the HI lines are
predicted to be much narrower than the observed ones, i.e. FWHMs of the order
of 100 \kms would be obtained, in contrast with our observations. 

In conclusion, our analysis 
shows that neither a spherically symmetric wind nor spherically
symmetric accretion can reproduce the Brackett and Paschen line ratios
and fluxes.
There are a few recent observational results suggesting that 
both the emission from the accretion columns and from a
wind may contribute to the line profiles.
\citet{takami01} and  \citet{whelan04}, for example, found that the gas 
from a wind contributes to the emission in the wings of their \Ha\, and
\pab\, lines, thus producing line profiles broader than the ones predicted by
magnetospheric accretion models.
In our case, however, the observed H lines do not show the 
core/wings structure which is expected when both the accreting material and
a wind are contributing to the emission.

To investigate further the possibility of a co-existence of emission 
from both inflowing and outflowing gas in our H lines we analysed critically
the spectro-astrometric results as described below. 

\subsection{Insights from spectro-astrometry}
\label{sect:spectro-astrometry_disc}

In the last two Sections we showed that both the accretion model and the 
wind model fail in reproducing all the characteristic of the observed NIR HI 
lines.
A powerful  technique  to investigate the origin of the HI emission and 
separate the accretion and the wind contributions to the line profiles is  
spectro-astrometry \citep[see, e.g.,][]{takami01,whelan04,whelan05}.   
In fact, outflowing gas presents an extended emission testified by a detectable
positional shift, while accreting gas remain concentrated on the source
location. 

Although \citet{takami01} detected extended emission in the wings of the \Ha\, 
line up to $\sim$3 AU from RU Lupi, we found that 
there is no spectro-astrometric signal in the HI NIR lines 
(see Fig.~\ref{spectro_astrometry}). The lack of signal indicates that: 
(i) either the emission comes from a compact region 
(smaller than $\sim$1.3 AU);
(ii) or, if there is a further emission from an extended region,
this is symmetric, thus not detectable through spectro-astrometry, 
or too faint with respect to the emission from the star and the accretion
columns (see below).  

On the contrary, we do detect after subtraction of possible biases a 
substantial spectro-astrometric signal in the HeI \lam10830 line 
(Fig.~\ref{spectro_astrometry}, panels C, C').  
The shift found in the HeI line 
is interpreted as an emission originating in the  base of 
the  jet, that  partially fills the  absorption feature  of the  
``P Cygni'' profile due to the inner wind.
The inner  stellar  wind acts  like a  natural
coronograph   hiding  the  star   along  the   blueshifted  absorption,
increasing  the contrast of emission from outer regions.
This could explain why we detect a positional shift in the HeI line and
not in the NIR HI lines.
A similar extended emission superimposed to the blueshifted
absorption feature of the HeI line was found by \citet{takami02} in DG Tauri. 
Following the same reasoning of \citet{takami02} we suggest that there are 
three contributes to the profile of the HeI line:
(i) red-shifted emission presumably due to accreting gas;
(ii) blue-shifted absorption due to the presence of an inner spherical wind 
emerging from the stellar surface;
(iii) blue-shifted extended emission 
between V$\sim$-50 \kms\, and V$\sim$-200 \kms\, 
superimposed to the absorption feature and
coming from a jet at a larger distance from the source. 
This emission is supposed to come from the base of the
collimated micro-jet detected by \citet{takami01} in the forbidden
\s\, and \oi\, lines and in the \Ha\, line. 
The velocity range covered by the HeI extended component is the same as in
the \Ha\, line.
The HeI spectro-astrometric signal extends up to $\sim$20 mas from the source 
($\sim$3 AU). However, as explained in Sect.~\ref{sect:spectro_astrometry}, 
this is only a lower limit to the true angular scale
of the emission region which could be more extended.

Combining in a bi-dimensional
plane the shifts detected along the parallel and the perpendicular slits we
find that the extended emission is not completely aligned with the
jet of  \citet{takami01}, the misalignment being not greater than 30\degr.
This may be justified by the fact that:
(i) the jet may be not well collimated in the He line;
(ii) our  observations were made 4 years  later and  a 200~\kms\, stream
moves  1~AU  in  a   week.   Furthermore,  \citet{takami01}  do  observe
variability  in  their spectro-astrometric  data  taken several  years
apart. 
Another discrepancy with respect to the result obtained by \citet{takami01}
in the H$\alpha$ line is that we did not detect extended emission in the 
red-shifted part of the HeI line.
This can be due, again, to the fact that the jet emission in the HI and HeI
lines is fainter than in the optical H$\alpha$ line and it can be evidenced
by spectro-astrometry only over the blueshifted absorption, which hides the
emission from the star and increases the contrast of outer emission.


\section{Conclusions}
\label{sect:conclusions}

In this paper we analysed infrared spectra and images of the T Tauri star 
RU Lupi, in order to find observational constraints for the physics of the
circumstellar region.
This source was previously observed in the optical wavelength range by 
other authors, that reported about a strong accretion activity  and 
the presence of a micro-jet detected in the optical
forbidden lines and in the H$\alpha$ line \citep{takami01}. 
We have observed for the first time RU Lupi in the NIR with high angular
and spectral resolution.
In this paper we report about these observations and, in particular, 
we investigate the origin of the permitted H and He NIR
lines through the analysis of the line profiles and fluxes, and using the 
spectro-astrometry technique which allows us to constrain the emission
region of each velocity component of the detected lines.

The problem of the origin of HI lines was widely debated
in the last decades.
Many different models were proposed to reproduce their profiles and fluxes:
wind models \citep[e.g., ][]{hartmann90}, accretion models 
\citep[e.g., ][]{hartmann94,muzerolle98a,muzerolle01} and also hybrid models
accounting for the emission from both the accretion columns and the wind
\citep[e.g., ][]{kurosawa06}.
These analyses, however, focused mainly on Balmer lines observed in the
optical range.
The Paschen and Brackett lines detected in our NIR spectra of RU Lupi
present broad, slightly blueshifted and nearly symmetric profiles, which are
difficult to reproduce with either a wind or an accretion model.
This was already noted, in general,
 by \citet{folha01}, \citet{whelan04}, \citet{nisini04}.
We tentatively used a toy model of a spherical envelope of partially
ionised Hydrogen with a  wind or an accretion velocity profile 
\citep{nisini04} to constrain the HI line ratios and fluxes.
This analysis showed that neither a spherical wind nor spherical
accretion can reproduce the Brackett and Paschen decrement
and fluxes.
 
Our spectro-astrometric analysis did not highlight any extended
emission in the Paschen and Brackett lines, suggesting that the region 
emitting the HI NIR lines is very compact ($<$1.3 AU) and/or symmetric. 
On the other hand, the HeI \lam10830 line blueshifted absorption feature 
clearly indicates the presence of an inner wind.
The wide range of velocities covered by the absorption feature 
favours the geometry
of a spherical wind emerging from the stellar surface rather than a disk
wind as found in many TTS by \citet{edwards06}.
In addition, the spectro-astrometric analysis highlighted the presence 
of an emission superimposed to the absorption feature that extends up 
to at least $\sim$3 AU from the source and covers the same velocity range
of the H$\alpha$ extended emission found by \citet{takami01}.
We suggest that this emission comes from the blue lobe of the 
micro-jet detected in the optical lines by \citet{takami01}. 
Interestingly, the HeI showed to be more sensitive than the HI NIR lines 
to faint extended emission because the absorption feature increases 
the contrast between the emission from the extended region and the
continuum from the source.
This confirms the potential of the HeI \lam10830 line to investigate
the complex geometry of the inner part of the circumstellar region of CTTS,
where both the accretion and the ejection processes take place
\citep[see also ][]{edwards06}. 
On the base of these results, future high angular resolution observations of
the permitted H and He lines in CTTS can give useful hints to the 
understanding of the accretion/ejection mechanisms and the development of 
theoretical models accounting for the emission from both the outflowing and
the accreting gas.

\begin{acknowledgements}
Linda Podio thanks CAUP for hospitality during the development of this work. 
We are also grateful to the referee, Dr. Suzan Edwards, for the attentive comments
which allowed us to improve the first version of this paper.
This work was partially supported by the European Community's Marie Curie 
Research and Training Network JETSET (Jet Simulations, 
Experiments and Theory) under contract MRTN-CT-2004-005592, 
and by the Consiglio Nazionale delle Ricerche in the framework of a 
CNR/GRICES Agreement. 
P.J.V.G. work was supported in part by the Funda\c{c}\~ao para a  Ci\^encia e 
a  Tecnologia through  projects POCI/CTE-AST/55691/2004 and 
PTDC/CTE-AST/65971/2006 from POCTI, with funds  from the European 
program FEDER. 
\end{acknowledgements}

\end{document}